\documentclass[aps,prl,twocolumn,superscriptaddress,showpacs]{revtex4}
\usepackage{graphicx}
\usepackage{mathrsfs}

\begin{document}



\title{Classical-to-stochastic Coulomb blockade cross-over in aluminum arsenide wires}


\author{J.~Moser}
\email{moser@wsi.tu-muenchen.de}
\affiliation{Walter Schottky Institut, Technische Universit\"at
M\"unchen, D-85748 Garching, Germany\\}
\affiliation{NEST-CNR-INFM, Scuola Normale Superiore, I-56126 Pisa, Italy\\}

\author{S.~Roddaro}
\affiliation{NEST-CNR-INFM, Scuola Normale Superiore, I-56126 Pisa, Italy\\}

\author{D.~Schuh}
\affiliation{Walter Schottky Institut, Technische Universit\"at
M\"unchen, D-85748 Garching, Germany\\}

\author{M.~Bichler}
\affiliation{Walter Schottky Institut, Technische Universit\"at
M\"unchen, D-85748 Garching, Germany\\}

\author{V.~Pellegrini}
\affiliation{NEST-CNR-INFM, Scuola Normale Superiore, I-56126 Pisa, Italy\\}

\author{M.~Grayson}
\affiliation{Walter Schottky Institut, Technische Universit\"at
M\"unchen, D-85748 Garching, Germany\\}


\date{13 December 2005}

\begin{abstract}

We report low-temperature differential conductance measurements in aluminum arsenide cleaved-edge overgrown quantum wires in the pinch-off regime. At zero source-drain bias we observe Coulomb blockade conductance resonances that become vanishingly small as the temperature is lowered below $250\,{\rm mK}$. We show that this behavior can be interpreted as a classical-to-stochastic Coulomb blockade cross-over in a series of asymmetric quantum dots, and offer a quantitative analysis of the temperature-dependence of the resonances lineshape. The conductance behavior at large source-drain bias is suggestive of the charge density wave conduction expected for a chain of quantum dots.

\end{abstract}

\pacs{73.21.Hb,73.23.-b,73.23.Hk,73.20.Mf}

\maketitle



Since the observation of Coulomb blockade (CB) in disordered micron-long GaAs wires~\cite{field}, electrons in GaAs have been the preferred system of CB study due to their small mass and long mean-free-path, allowing the necessary control of quantum confinement length scales below the mean free path to design shorter, single-resonance systems.  However the original literature~\cite{field} also includes intriguing results in heavier-electron mass disordered Si wires which have proven harder to interpret, presumably because they contain a multiplicity of resonances.  Like the GaAs CB, the Si system shows a suppressed conductance for a range of gate voltage ($V_{GD}$, drain $D$ grounded) and source-drain voltage ($V_{SD}$) which periodically closes at a resonance forming a diamond-shaped conductance suppression in $V_{SD}$ {\sl vs.} $V_{GD}$.  However unlike GaAs, the resonance peaks in Si wires are very small, and at finite $V_{SD}$ bias outside of the diamond region exhibit a comparatively large single peak followed by a continuous tail at larger $V_{SD}$. Theories have been proposed to explain this behavior \cite{ruzin,likharev}, but a quantitative confirmation which extracts the model parameters from the data has been lacking.  More recently, experiments on split-gate constrictions in the quantum Hall (QH) regime have shown finite bias conductance traces strikingly similar to that of the disordered Si wire literature~\cite{roddaro}.  This behavior is linked to forward and reverse moving QH edge modes coupled via an interaction region artificially induced by the constriction.  Although the one-dimensional (1D) system is comprised of GaAs electrons, the soft confining potential of the split-gate region is expected to slow the mode velocity, effectively increasing the transport mass in this system.

\begin{figure}[t]
\includegraphics[width=8.5cm]{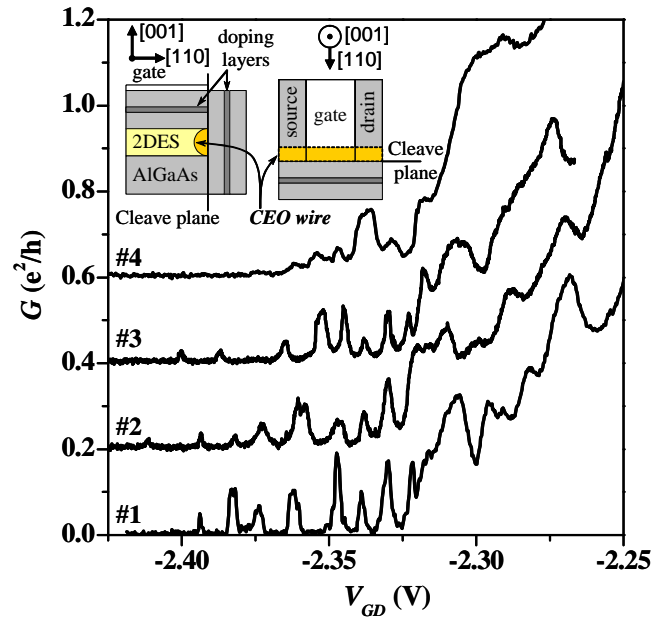}
\caption{(Color) Conductance $G$ {\sl vs.} gate bias $V_{GD}$ for several $V_{GD}$ sweeps in the positive-voltage direction at a base electron temperature $T=100\,{\rm mK}$. Beyond the fourth sweep it becomes increasingly difficult to resolve resonances. Each trace is offset by $0.2 e^2/h$ for clarity. Inset: schematic of the sample.}
\label{resonance_combs}
\end{figure}

In this Letter, we present a disordered heavy-mass AlAs wire system in the Coulomb blockade regime.  Similar features to the above systems are observed and studied in detail.  We propose that this behavior is universal to heavy-mass disordered 1D systems at pinchoff, and we illustrate how quantitative information about the relevant capacitances in the problem can be determined.  We find that the transport theories proposed by Ruzin, {\sl et al.}~\cite{ruzin} and Bakhvalov, {\sl et al.}~\cite{likharev} for a series of weakly coupled dots are able to explain the CB resonances and the CB conductance outside the diamond, respectively.  As with standard CB, the shape of the diamond allows us to deduce the capacitance of the principle dot in the wire, but we then proceed, according to Ref.~\cite{ruzin}, to fit the temperature dependence of the resonance to a classical-to-stochastic CB crossover which allows us to determine the capacitance of the second limiting dot.  Large bias conductances outside the Coulomb diamond are then shown to be consistent with a charge density flowing through a chain of dots described in Ref.~\cite{likharev} pinned at low energies by the CB gap.

The quantum wire samples are fabricated from a $150 {\rm \AA}$-wide, modulation-doped AlAs 2D electron system (2DES) sandwiched between two AlGaAs spacers and grown on a (001) GaAs substrate (see inset to Fig.~1). The substrate is cleaved {\it in-situ} at the perpendicular (110) plane and overgrown with another modulation-doped barrier~\cite{moser}. Electrons in the two degenerate X-valleys accumulate along the cleaved edge of the 2DES and form the doubly valley-degenerate 1D system. A $1\mu{\rm m}$-wide metal gate on the substrate depletes the 2DES underneath and varies the wire density in the segment located below it. The 2DES regions on each side of the gate serve as ohmic contacts, and conductance measurements are performed in a 2-point configuration using $V_{ac} \simeq 10 \mu V < k_{B}T$ and standard lock-in techniques.  The effective mass in the direction of the wire is ${\rm m}^{*}=0.33{\rm m}_{0}$ in units of the free-electron mass, which is a factor of 5 larger than in CEO GaAs wires~\cite{yacoby_ssc}, leading to a stronger role for interactions at the same density.  Samples are cooled down in a dilution refrigerator with an electron base temperature $T = 100\,{\rm mK}$. The substrate and the cleaved-edge are illuminated at $10\,{\rm K}$ with two infrared light-emitting diodes.

\begin{figure}[t]
\includegraphics[width=8.5cm]{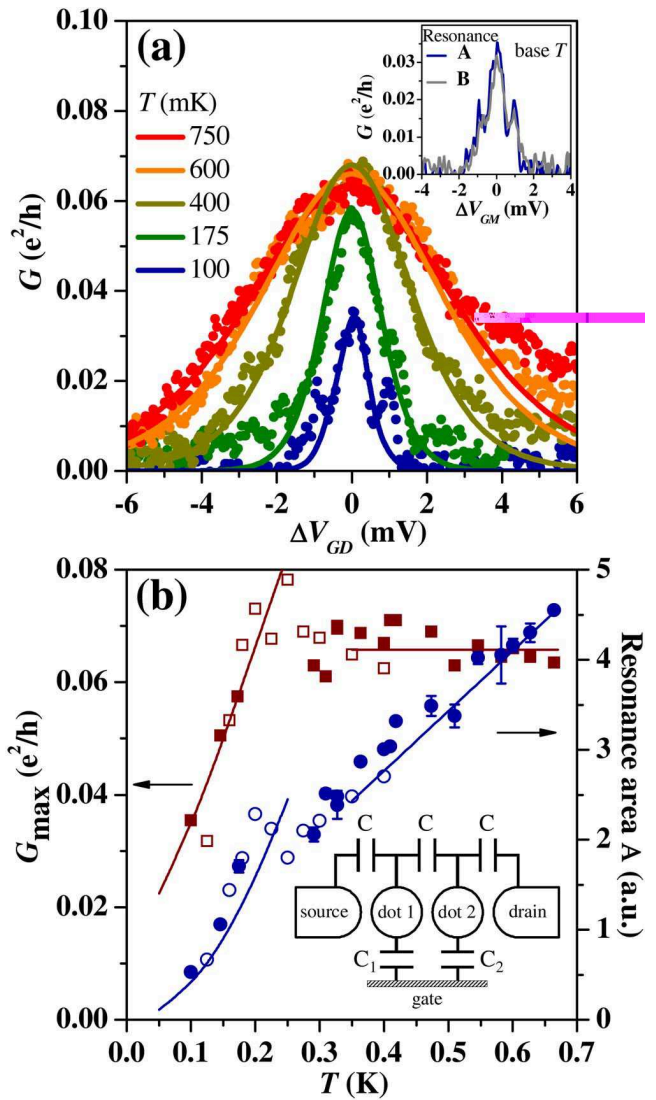}
\caption{(Color) (a) Evolution of a resonance {\bf A} with temperature $T$. Curve fits of the model of Ref.~\cite{ruzin} for two asymmetric dots in series are shown assuming capacitances described in the text. Resonances {\bf A} and {\bf B} are shown in the inset to share a similar triple-peak structure at base $T$. (b) Conductance peak $G_{max}$ (red) and resonance area $\mathscr{A}$ (blue) as a function of $T$. Filled (hollow) symbols correspond to resonance {\bf A} ({\bf B}).  $\mathscr{A}$ is twice the integrated area of the left half of the resonance; error bars quantify the left-right asymmetry at higher temperatures, and correspond to $\pm \left|\mathscr{A} - \text{full area} \right|/2$. Lines are fit to the model of Ref.~\cite{ruzin}.}
\label{resonance_T}
\end{figure}

Figure~1 depicts combs of conductance resonances at $T=100\,{\rm mK}$ as the gate bias $V_{GD}$ is scanned past the pinch-off threshold. The first $V_{GD}$ sweep after illumination and at base $T$ yields quasi-periodic resonances.  Upon sweeping $V_{GD}$ repeatedly in the same direction some resonances appear to be quite robust while others shift. By the fourth gate sweep most resonances have vanished and those remaining are lumped close to threshold.

We interpret this behavior as Coulomb blockade (CB) in a disorder potential in analogy with its original discovery in disordered Si wires~\cite{field}. As the wire is depleted, the disorder potential isolates 0D islands of electrons throughout the wire, whose capacitance defines a charging energy which periodically blocks single electron transport.  Presumably, cycling $V_{GD}$ affects the distribution of ionized dopants underneath the gate~\cite{davis}, inducing variations in the potential background seen by the wire at each cycle. However by narrowing the gate bias window it is possible to preserve one resonance upon multiple $V_{GD}$ cycles and perform a systematic study of it, as we now demonstrate.

We study two resonances {\bf A} and {\bf B} in detail. Each is observed in a separate cooldown, and is well isolated and located at a similar $V_{GD}$ away from threshold.  Fig.~2a illustrates the influence of $T$ on resonance {\bf A}, and Fig.~2b summarizes the $T$-dependence of the peak conductance $G_{max}$ and of the resonance area $\mathscr{A}$ for {\bf A} and {\bf B}. Two distinct temperature regimes are apparent in Fig.~2b: $G_{max}$ is weakly $T$-dependent above $250\,{\rm mK}$, then sharply falls off at lower $T$. $\mathscr{A}(T)$ is linear for $T>250\,{\rm mK}$ but drops rapidly for $T < 250\,{\rm mK}$.  At base $T$ both resonances resolve into a striking triple-peak (inset to Fig.~2a).

The $T$-dependence of the resonance lineshape in Fig.~2a suggests that transport within our wire is limited by two asymmetric quantum dots. This regime has been addressed theoretically by Ruzin, {\sl et al.}~\cite{ruzin} for the case of two quantum dots in series, wherein tunneling is incoherent and density is tuned by a common gate (see inset to Fig.~2b).  Following the notation of Ref.~\cite{ruzin}, individual dots 1 and 2 are characterized by their capacitances to the gate $C_{1}$ and $C_{2}$. The inter-dot capacitance as well as the capacitance between each dot and either source or drain is assumed to be equal to $C$.  The total capacitance of dot $i$ is $C_i^{tot}=C_i+2C$.  Transport through the double dot structure occurs only when both dots have an available energy level within $k_BT$ of the Fermi energy. In the case of strongly asymmetric dots, {\sl e.g.} $C_{1}^{tot} \ll C_{2}^{tot}$, the large difference in the level spacings $e^{2}/C_{1}^{tot} \gg e^{2}/C_{2}^{tot}$ makes this condition increasingly difficult to meet at low $T$, leading to stochastic CB and vanishing resonances. On the other hand, for $k_{B}T > e^{2}/C_{2}^{tot}$, the model in Ref.~\cite{ruzin} explains that the level spacing in dot 2 is not relevant and transport is dominated by the charging of the single dot 1. Indeed, looking at Fig.~2b, $G_{max}(T)$ and $\mathscr{A}(T)$ {\sl above} $250\,{\rm mK}$ resemble the behavior of a single dot resonance in the classical CB regime~\cite{fwhm}.

We therefore posit $e^{2}/C_{2}^{tot}=k_{B}\times 250\,{\rm mK}=21 \mu{\rm eV}$ ($C_2^{tot} = 7.6\,{\rm fF}$) as the cross-over energy scale below which the double dot structure becomes relevant \cite{ruzin}. Given that the periodicity of CB oscillations is determined by the smallest gate-dot capacitance $C_{1}\ll C_{2}$, we use the energy conversion $\partial E_{F}/\partial V_{GD} = 40\,{\rm meV/V}$ measured from the thermal linewidth~\cite{fwhm}, along with the typical resonance spacing $\delta V_{GD} \simeq 10\,{\rm mV}$ \footnote{This value was measured as the distance between resonance {\bf A} and its immediate neighbor. This also agrees with data in Fig.~1.}, to obtain $e^{2}/C_{1}^{tot}\simeq 400 \mu{\rm eV}$ ($C_1^{tot} = 0.4\,{\rm fF}$). The resonant conductance through the double dot can now be calculated following the model in Ref.~\cite{ruzin} and using the estimated $C_{1}^{tot}/C_{2}^{tot}$ ratio. The lineshape evolution is captured by the simple analytical solution obtained in the limit of small interdot capacitance $C$.  The complete set of calculated resonance lineshapes are plotted in Fig.~2a, with the solid lines in Fig.~2b representing the calculated peak conductances and areas.

\begin{figure}[t]
\includegraphics[width=8.5cm]{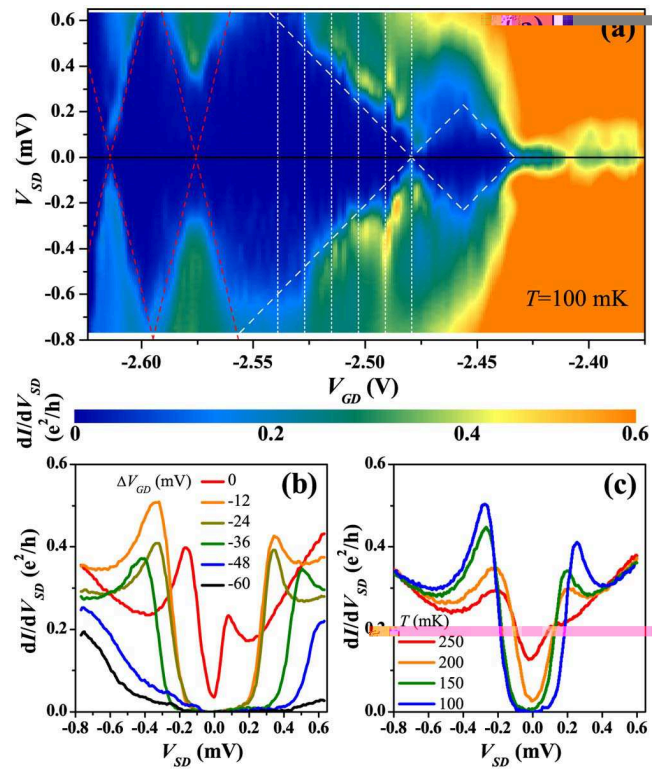}
\caption{(Color) (a) Differential conductance $dI/dV_{SD}$ as a function of source-drain dc bias $V_{SD}$ and gate bias $V_{GD}$ at $T=100\,{\rm mK}$. Diamonds are highlighted. (b) Subset showing traces at fixed $V_{GD}$ indicated by the vertical dashed lines in (a). The conductance at $\Delta V_{GD} = 0$ and $V_{SD}=0$ corresponds to the maximum of a resonance peak. (c) Temperature dependence of $dI/dV_{SD}$ at fixed $V_{GD}\sim 50\,{\rm mV}$ from pinch-off, taken from a different dataset.}
\label{spectroscopy}
\end{figure}

The gate-dot capacitance of the current-limiting dot is $C_1=e/\delta V_{GD}\simeq0.016\,{\rm fF}$ \footnote{The dot length $L$ can be extracted \cite{defranceschi} by fitting $C_{1}$ to an electrostatic model of a cylinder-shaped dot $C_1\simeq 2\pi \epsilon \mathrm{ln}(2d/R)\times L$, where $\epsilon$ is the absolute dielectric constant of AlGaAs, $d=380\,{\rm nm}$ is the gate-dot distance and $R=7.5\,{\rm nm}$ is the dot radius, yielding $L \simeq 100\,{\rm nm}$.}.  $C_1$ is small compared to $C_1^{tot}$ so that the finite coupling capacity $C$ is expected to play a role. Indeed, after Refs.~\cite{ruzin,matveev}, a finite $C$ does not change the overall $T$-dependence of $G_{max}$ and $\mathscr{A}$ but could explain the emergence of finer features at low temperature, such as the multiple peaks shown in the inset to Fig.~2a.

Having demonstrated evidence of a classical-to-stochastic CB crossover in AlAs wires, we now focus on the source-drain bias dependence and analyze the non-linear conduction through the wire. The color plot in Fig.~3a reports the differential conductance $dI/dV_{SD}$ as a function of $V_{GD}$ and $V_{SD}$ at $T=100\,{\rm mK}$. Data are collected by sweeping $V_{SD}$ and stepping $V_{GD}$. All features evolve smoothly, without any significant trapped charge events. CB diamonds defined as regions of vanishing conductance are highlighted with dashed lines.  Note that the vanishingly small resonances at $V_{SD}=0$ are not visible on this color scale.  Near threshold, CB diamonds highlighted in white are visible. Additional diamonds with a steeper $dV_{GD}/dV_{SD}$ slope are observed at more negative $V_{GD}$ values (red dashed lines). Fig.~3b shows several $dI/dV_{SD}$ {\sl vs.} $V_{SD}$ traces at fixed $V_{GD}$ as indicated by the vertical dashed lines in Fig.~3a: a gap structure centered at $V_{SD}=0$ develops beyond pinch-off as $T$ is lowered (Fig.~3c). Similar non-linear features are observed for three separate cooldowns of the same sample (data not shown). Note how the condition of maximum resonant conduction at the center of the CB diamond $\Delta V_{GD} = 0$ in Fig.~3b corresponds to a strong minimum in $dI/dV$ {\sl vs.} $V_{SD}$ similar to the observations in Ref.~\cite{field}. The dot size for the cooldown of Fig. 3 can be estimated, as before, from $\delta V_{GD}\simeq38\,{\rm mV}$, yielding $L\simeq30\,{\rm nm}$. Comparing with the estimate for $L$ obtained from Fig.~1 indicates that the dot size fluctuates with cooldown. In addition, an analysis of the red and white diamond slopes shows that the limiting dot becomes more and more isolated as the electron density decreases.

\begin{figure}[t]
\includegraphics[width=8.5cm]{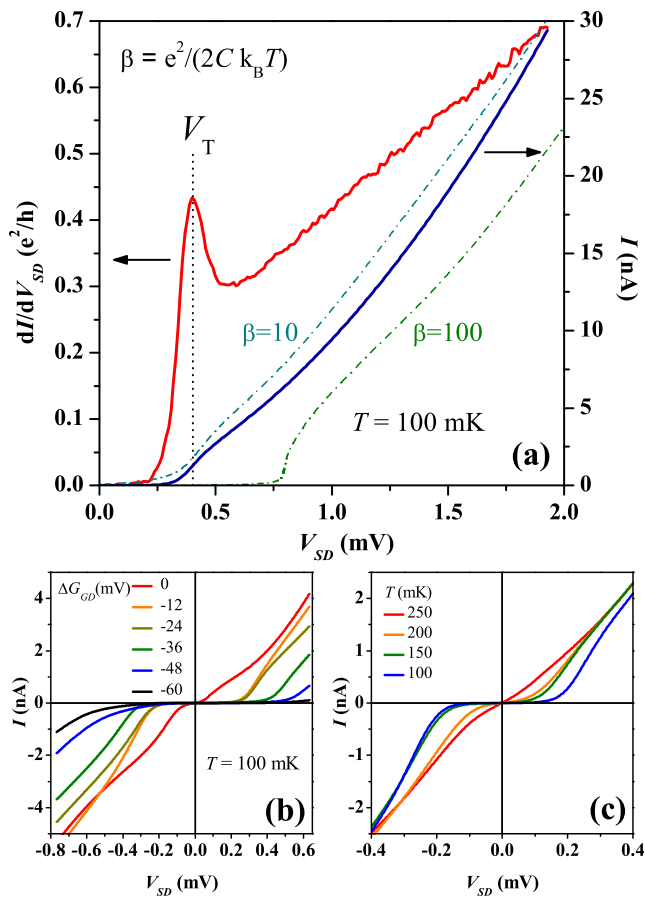}
\caption{(Color) (a) Typical $dI/dV_{SD}(V_{SD})$ and integrated current $I(V_{SD})$. $\beta$ is ratio of the charging energy over the thermal energy. (b) Integrated $I(V_{SD})$ from traces of Fig.~3b. (c) Temperature dependence of $I(V_{SD})$ after integration of $dI/dV_{SD}$ in Fig.~3c.}
\label{cdw}
\end{figure}

The gap structure in the differential conductance $dI/dV_{SD}(V_{SD})$ reported in Fig.~3 displays peculiar features. Outside of the CB gap the conductance of Fig.~3b shows a single peak followed by a monotonically increasing conductance.  This has to be contrasted with standard CB behavior which would show a series of peaks outside the gap region.  This first peak also shows the temperature broadening expected for CB features (Fig.~3c), but the wings at higher voltage show no obvious $T$-dependence.

This conductance threshold outside of the CB gap is reminiscent of what one would expect for a de-pinned charge density wave, as first proposed in such systems in Ref.~\cite{field}.  This effect is more easily recognized if the $dI/dV_{SD}$ data are integrated to give $I-V_{SD}$.  Fig.~4a shows an example of a $dI/dV_{SD}$ curve integrated to reveal a step-like current onset in $I-V_{SD}$ outside of the CB region.  The full dataset of Figs.~3b and 3c are integrated for comparison in Figs.~4b and 4c.  We can now identify the peak in $dI/dV_{SD}$ in Fig.~4a as a threshold voltage $V_{T}$ at which point the current is at the center of its step-like rise, and the charge flow becomes de-pinned. Ref.~\cite{likharev} provides a quantitative description of how a chain of coupled dots in series can support current flow in the form of charge solitons, so for comparison we plot the two theoretical curves available from Ref.~\cite{likharev} in Fig.~4a along with the experimental data.  These curves represent the cases $\beta=e^{2}/(2Ck_{B}T)=10$ and 100, appropriate to the estimated capacitance $C=0.2\,{\rm fF}$ that yields $\beta \simeq 45$ at $T=100\,{\rm mK}$.  Ref.~\cite{likharev} also predicts that the conductance threshold $V_{T}(V_{GD})$ resembles Coulomb diamonds that do not close, as observed here in experiment.  We propose that flow of charge density through the chain of quantum dots is pinned by the CB gap of the dot with the smallest capacitance, and outside of this gap the charge flows through the dot chain as a soliton waves as predicted for such systems.

We note that different $T$-dependent resonance areas were recently measured in another 1D experiment by Auslaender, {\sl et al.} \cite{auslaender}, and were interpreted as evidence for Luttinger liquid physics. Yet the fact that the dot-chain model successfully explains our measurements suggests that our disordered quantum wire is in a qualitatively different conductance regime.

In summary, we have provided evidence that disordered AlAs CEO wires break up into a series of weakly-coupled quantum dots at low electron density. The temperature dependence of resonances is interpreted as a classical-to-stochastic Coulomb blockade cross-over dominated by two asymmetric dots. Differential conductance measurements indicate a threshold for conductance outside the Coulomb diamond which can be explained with the soliton model of conductance in a chain of coupled quantum dots.  The analysis provided here should prove useful for identifying transport mechanisms in similar large-mass, one-dimensional disordered systems.

$\\$
We thank J. Weis, T. Giamarchi, Y. Meir, D. Goldhaber-Gordon and G. Abstreiter for stimulating discussions. We gratefully acknowledge support from the COLLECT EC-Research Training Network, HPRN-CT-2002-00291, German BmBF grant 01 BM 470, and FIRB "Nanoelettronica".


\begin{references}

\bibitem{field}
S. B. Field, M. A. Kastner, U. Meirav, J. H. F. Scott-Thomas, D. A. Antoniadis, H. I. Smith, and S. J. Wind, Phys. Rev. B {\bf 42}, 3523 (1990).

\bibitem{ruzin}
I. M. Ruzin, V. Chandrasekhar, E. I. Levin, and L. I. Glazman, Phys. Rev. B {\bf 45}, 13469 (1992).

\bibitem{likharev}
N. S. Bakhvalov, G. S. Kazacha, K. K. Likharev, and S. I. Serdyukova, Soviet Physics - JETP {\bf 68}, 581 (1989).

\bibitem{roddaro}
S. Roddaro, V. Pellegrini, F. Beltram, L. N. Pfeiffer, and K. W. West, Phys. Rev. Lett. {\bf 95}, 156804 (2005).

\bibitem{moser}
J. Moser, T. Zibold, S. Roddaro, D. Schuh, M. Bichler, F. Ertl, G. Abstreiter, V. Pellegrini, and M. Grayson, Appl. Phys. Lett. {\bf 87}, 052101 (2005).

\bibitem{yacoby_ssc}
A. Yacoby, H. L. Stormer, K. W. Baldwin, L. N. Pfeiffer, and K. W. West, Solid State Commun. {\bf 101}, 77 (1997).

\bibitem{davis}
J. H. Davis, Semicond. Sci. Technol. {\bf 3}, 995 (1988).

\bibitem{fwhm}
I. O. Kulik, and R. I. Shekhter, Zh. Eksp. Teor. Fiz. {\bf 68}, 623 (1975) [Sov. Phys. JETP {\bf 41}, 308 (1975)].

\bibitem{matveev}
K. A. Matveev, L. I. Glazman, and H. U. Baranger, Phys. Rev. B {\bf 54}, 5637 (1996).

\bibitem{delsing}
L. S. Kuzmin, P. Delsing, T. Claeson, and K. K. Likharev, Phys. Rev. Lett. {\bf 62}, 2539 (1989).

\bibitem{weis}
J. Weis, in {\sl Functional Nanostructures}, Lect. Notes Phys. {\bf 658} (2004).

\bibitem{defranceschi}
S. De Franceschi, J. A. van Dam, E. P. A. M. Bakkers, L. F. Feiner, L. Gurevich, and L. P. Kouwenhoven, Appl. Phys. Lett. {\bf 83}, 344 (2003).

\bibitem{auslaender}
O. M. Auslaender, A. Yacoby, R. de Picciotto, K. W. Baldwin, L. N. Pfeiffer, and K. W. West, Phys. Rev. Lett. {\bf 84}, 001764 (2000).

\end{references}
\end{document}